\documentclass[12pt]{iopart}
\newcommand{\rn}{\mathcal{R}_n}
\newcommand{\ff}{\delta^2 F_{\rn} - n \delta^2 F}

\newcommand{\cp}{\mathbb{C}^+}
\usepackage{color}
\definecolor{Blue}{rgb}{0.3,0.3,0.9}
\usepackage{iopams}  
\begin{document}

\title[Corrections to scaling in entanglement entropy from boundary perturbations]{Corrections to scaling in entanglement entropy from boundary perturbations}

\author{Erik Eriksson and Henrik Johannesson}

\address{Department of Physics, University of Gothenburg, SE 412 96 Gothenburg,
Sweden}
\ead{erik.eriksson@physics.gu.se and henrik.johannesson@physics.gu.se}
\begin{abstract}
We investigate the corrections to scaling of the R{\'e}nyi entropies of a region of size $\ell$ at the end of a semi-infinite one-dimensional system described by a conformal field theory when the corrections come from irrelevant boundary operators. The corrections from irrelevant bulk operators with scaling dimension $x$ have been studied by Cardy and Calabrese (2010), and they found not only the expected corrections of the form $\ell^{4-2x}$ but also unusual corrections that could not have been anticipated by finite-size scaling arguments alone. However, for the case of perturbations from irrelevant \textit{boundary} operators we find that the only corrections that can occur to leading order are of the form $\ell^{2-2x_b}$ for boundary operators with scaling dimension $x_b < 3/2$, and $\ell^{-1}$ when $x_b > 3/2$. When $x_b=3/2$ they are of the form $\ell^{-1}\log \ell$. A marginally irrelevant boundary perturbation will give leading corrections going as $(\log \ell )^{-3}$. No unusual corrections occur when perturbing with a boundary operator.
\end{abstract}

\maketitle

\section{Introduction}

The block entanglement of a quantum system has been found to be a powerful tool for characterizing the scaling behavior near a quantum critical point \cite{CC2}. For a system in a pure state and with the Hilbert space partitioned into a direct product ${\cal H}={\cal H}_{A} \otimes {\cal H}_{B}$ (with A and B the corresponding two parts of the system), the block entanglement is encoded by the von Neumann entropy $S_{A} = - \mbox{Tr}\rho_{A} \log \rho_{A}$ of the reduced density matrix $\rho_{A}$, with $S_{A} = S_{B}$. The most interesting case is in one dimension. For an infinite system with an interval A of length $\ell$ the asymptotic behavior of the von Neumann entropy is given by  \cite{CC1}
\begin{equation} \label{vonNeumannScaling}
S_{A}  \sim \frac{c}{3}\log \frac{\ell}{\epsilon} + c_1^{\prime}
\end{equation}
near the critical point. Here $c$ is the central charge of the underlying conformal field theory. The constant $\epsilon$ is an arbitrary cutoff scale, with $c_1^{\prime}$ also being a non-universal number. 
As a way to characterize the full entanglement spectrum one may introduce an additional parameter $n$, with $n$ a positive real number, and define the R\'{e}nyi entropies
\begin{equation} \label{S_An}
S_{A}^{(n)}= \frac{1}{1-n} \log \textrm{Tr} \, \rho_{A}^{n} \, ,
\end{equation}
with $\lim_{n \rightarrow 1} S_{A}^{(n)}= S_{A}$.
As expected from finite-size scaling theory, the critical scaling $S_{A}^{(n)} \sim (c/6)(1+n^{-1})\log (\ell / \epsilon)$ of the R\'{e}nyi entropies exhibit $\mathcal{O}(\ell^{4-2x})$ corrections \cite{CC3}. Here $x > 2$ is the scaling dimension of the leading irrelevant operator (with ''irrelevant'' being understood in the sense of the renormalization group). As shown in Ref.~\cite{CC3}, there can also be unusual $n$-dependent corrections of $\mathcal{O}(\ell^{-2x/n})$ and $\mathcal{O}(\ell^{2-x-x/n})$, where, in the first case, $x$ may in fact be less than 2, corresponding to a scaling correction produced by a relevant operator. These unusual corrections often come with an oscillating prefactor, which however vanishes when $n \rightarrow 1$ in all known cases~\cite{calaPRL,CE}. For a semi-infinite system, with a conformally invariant boundary condition (CIBC), operators in the bulk may produce additional unusual scaling corrections $\ell^{-x/n}$ to the R\'{e}nyi entropies, on top of the ordinary $\mathcal{O}(\ell^{2-x})$ corrections with $x>2$ predicted by finite-size scaling \cite{CC3}. In contrast to the case of an infinite system, the oscillating prefactor that multiplies the leading unusual $\ell^{-x/n}$-correction does not vanish in the limit $n \rightarrow 1$. This unexpected feature was first observed in numerical work in Ref.~\cite{LSCA}, and recently derived analytically for the case of the XX-chain with open boundary conditions~\cite{FC}.

In this article we inquire about the scaling corrections to the critical R\'{e}nyi entropies of a semi-infinite one-dimensional system which are generated by irrelevant {\em boundary operators}. Recall that boundary operators arise in the operator product expansion (OPE) of a chiral operator with its mirror image across the boundary. More precisely, given a boundary conformal field theory (BCFT) defined on the complex half-plane $\{z=\tau +iy \mid y\ge 0\}$ with a CIBC at $y=0$, the OPE of a chiral operator $\phi (\tau,y)$ with its mirror image $\phi(\tau,-y)$ reads \cite{CL,DiehlDietrich}
\begin{equation}
\phi(\tau,y) \phi(\tau,-y) \sim \sum_j \frac{C_{\phi,j}}{(2 y)^{2 x_{\phi}-x_j}}\phi_j(\tau), \ \ \ y \rightarrow 0.
\end{equation}
Here $x_{\phi}$ is the scaling dimension of $\phi$, and $\phi_j$ are boundary operators of dimension $x_j$. Nonzero values of the expansion coefficients $C_{\phi,j}$ select those boundary operators which are consistent with the particular CIBC imposed at $y=0$. Knowing the boundary operator content associated with a system allows for a complete characterization of its {\em boundary critical behavior}, i.e. those terms in the critical scaling of observables contributed by the presence of the boundary. For a quantum theory, where $\tau$ is a Euclidean time, this allows for identifying the long-time (a.k.a. low-energy) asymptotic critical behavior of the system close to the boundary. BCFT has a manifold of applications, spanning from open-string theory (D branes) \cite{String} to the study of quantum quenches \cite{CC4}.  A particularly important class of applications is that of a quantum impurity interacting with an electron liquid, where at low energies the impurity can be traded for a CIBC at the site of the impurity \cite{Affleck}. The increase of the block entanglement at quantum criticality due to the presence of the impurity is a universal number (\textit{boundary entropy}) which characterizes the type of boundary critical behavior. However, for a finite block there will always be additive corrections to the boundary entropy coming from irrelevant bulk {\em and} boundary operators. These corrections are expected to reveal features about quantum impurity phenomena which are otherwise difficult to access, the extent and character of the enigmatic "screening cloud" being a case in point \cite{ALS}.

\section{Scaling corrections from bulk operators: a brief review}

Consider a one-dimensional system with a boundary at $y = 0$ that is described by a BCFT. Let subsystem A be the region $0 \leq y \leq \ell$ and B the rest of the system, $y > \ell$. As shown by Calabrese and Cardy \cite{CC1,CC2}, $\mathrm{Tr} \, \rho_A^n$ (which enters the definition of the R\'{e}nyi entropies in Eq.~(\ref{S_An})) can be viewed as a path integral $Z_{\mathcal{R}_n}$ on an $n$-sheeted Riemann surface $\mathcal{R}_n$ with a boundary, and with proper normalization. Then 
\begin{equation}
\label{S_A2}
 S_A^{(n)}= \frac{1}{1-n} \log \frac{Z_{\mathcal{R}_n}}{Z^n} =  - \,\frac{\beta}{1-n}\,(F_{\mathcal{R}_n} - nF)\,   ,
\end{equation}
where $F = -\, \beta^{-1}\,$log$\,Z$ is the free energy, $F_{\rn} = -\, \beta^{-1}\,$log$\,Z_{\rn}$ and $\beta$ is the inverse temperature.
For an unperturbed BCFT, $Z_{\mathcal{R}_n} / Z^n$ can be calculated as a one-point function on the half-plane $\mathbb{C}^+$ of a \textit{twist field} $\Phi_n$ inserted at the branch point $z=$ i$\ell$ \cite{CC1}
\begin{equation} \label{twist}
\frac{Z_{\mathcal{R}_n}}{Z^n} = \langle \Phi_n(\mathrm{i}\ell)\rangle_{\mathbb{C}^+} = c_n \left(\frac{2\ell}{\epsilon} \right)^{-c(n-1/n)/12}\, ,
\end{equation}
where $\epsilon$ is the short-distance cutoff. This leads to 
\begin{equation} \label{S_A3}
S_A^{(n)} = \frac{c}{12} (1+\frac{1}{n}) \log \frac{2\ell}{\epsilon} + s^{A} + c'_n  \, ,
\end{equation}
where $s^{A} = \log g^{A}$ is the boundary entropy and $c'_n $ are non-universal constants \cite{CC1,ZBFS}. The well-known result for the critical block entanglement on a semi-infinite line,
\begin{equation} \label{S_A4}
S_A = \frac{c}{6} \log \frac{2\ell}{\epsilon} + s^{A} + c'  \, ,
\end{equation}
is obtained simply by letting $n \to 1^+$ in Eq.~(\ref{S_A3}).  

The corrections to this scaling behavior from irrelevant {\it bulk} operators was recently studied by Cardy and Calabrese~\cite{CC3}. Such a perturbation with a bulk operator $\Phi(z)$ having a scaling dimension $x > 2$ gives an action
\begin{equation} \label{action1}
S=S_{CFT} + \lambda \int\mathrm{d}^2 z \,\Phi(z)  \, ,
\end{equation}
where $\lambda$ is a coupling constant. The corrections to the free energies are given by the perturbation series
\begin{equation} \label{pert1}
F_{\mathcal{R}_n} = F_{\mathcal{R}_n}^{CFT} - {\displaystyle \sum_{N=1}^{\infty}} \,\frac{(-\lambda)^N}{N!} \, \int_{\mathcal{R}_n}\mathrm{d}^2 z_1  \cdots \int_{\mathcal{R}_n}\mathrm{d}^2 z_N  \, \langle \Phi(z_1) \cdots \Phi(z_N) \rangle_{\mathcal{R}_n} ,
\end{equation}
over the Riemann surface $\mathcal{R}_n$, and
\begin{equation} \label{pert2}
F = F^{CFT} - {\displaystyle \sum_{N=1}^{\infty}} \,\frac{(-\lambda)^N}{N!} \, \int_{\mathbb{C}^+}\mathrm{d}^2 w_1  \cdots \int_{\mathbb{C}^+}\mathrm{d}^2 w_N  \, \langle \Phi(w_1) \cdots \Phi(w_N) \rangle_{\mathbb{C}^+} ,
\end{equation}
over the ordinary complex half-plane $\mathbb{C}^+$. When the boundary conditions are such that $\langle \Phi(w) \rangle_{\mathbb{C}^+} \neq 0$ they found that the first-order correction to $F_{\mathcal{R}_n} - nF$ is of the form $\ell^{2- x}$. However, for $n > x / (x-2)$ they also found the appearance of the unusual $n$-dependent correction $\ell^{- x /n}$, that comes from the singularity at the branch point. For an infinite system without a boundary, the corrections take the forms $\ell^{4-2x}$ and $\ell^{-2x/n}$.

\section{Scaling corrections from irrelevant boundary operators}

We now wish to study the case when the perturbations come from irrelevant operators {\it on the boundary}. In doing so, we can follow the same procedure as Cardy and Calabrese in Ref.~\cite{CC3}. However, when perturbing with a boundary operator the surface integral of the perturbing field in the action (\ref{action1}) will be replaced by a line integral on the boundary. As we shall see, this will prevent the appearance of unusual $n$-dependent corrections. In fact, this is anticipated since $n$-dependent exponents only arise from the region at the branch point, which is located away from the boundary. Nevertheless, there still are results for the boundary case that do not follow from standard finite-size scaling analysis.

Thus, consider a BCFT on the upper half $y\geq 0$ of the complex plane $z=\tau + \mathrm{i}y$, so that $\tau$ is the boundary coordinate at $y=0$. The $n$-sheeted Riemann surface $\rn$ is then obtained by sewing together $n$ copies of this half-plane along $0 \leq y < \ell$ at $\tau = 0$. To evaluate the correlation functions on $\rn$ for a chiral operator $\phi(z)$ with scaling dimension $x_b$, we need to use the transformation property
\begin{equation} \label{trans}
\langle \phi(z_1) \cdots \phi(z_N) \rangle_{\mathcal{R}_n}  = \displaystyle \prod_{j=1}^{N} \left| \frac{\mathrm{d}z}{\mathrm{d}w}  \right|_{w=w_j}^{-x_b} \langle \phi(w_1) \cdots \phi(w_N) \rangle_{\mathbb{C^+}} \, ,
\end{equation}
where the map $z \mapsto w$ from $\rn$ to the upper half-plane $\mathbb{C}^+$ is given by
\begin{equation} \label{map}
w= -\mathrm{i}\, \frac{\left( \frac{z-\mathrm{i}\ell}{z+\mathrm{i}\ell} \right)^{1/n} +1}{\left( \frac{z-\mathrm{i}\ell}{z+\mathrm{i}\ell} \right)^{1/n} -1}\, .
\end{equation}
This gives
\begin{equation} \label{der}
\frac{\mathrm{d}z}{\mathrm{d}w} = -4 n \ell \, \frac{\left(\frac{w-\mathrm{i}}{w+\mathrm{i}} \right)^{n}}{(1+w^2)\left[\left(\frac{w-\mathrm{i}}{w+\mathrm{i}} \right)^{n}-1\right]^2}\, .
\end{equation}
Naturally, the mapping (\ref{map}) takes the boundary of $\rn$ to the boundary of $\cp$. Since $w$ is real on the boundary we see from (\ref{der}) that $|$d$z$/d$w|^{-x_b}$ is analytic on the boundary, as the only singularity is at $w=\,$i, i.e. when $z$ is at the branch point $z=\, $i$ \ell$. In particular, note that the point $|z|\to \infty$ gives a divergence in $|$d$z$/d$w|$ which only means that $|$d$z$/d$w|^{-x_b} \to 0$.

Now we can use this to study the scaling corrections of $S_A^{(n)} \propto (F_{\mathcal{R}_n} - nF)$ when adding a boundary perturbation,
\begin{equation} \label{action}
S=S_{CFT} + \lambda \int\mathrm{d} \tau \,\phi_b(\tau)  \, ,
\end{equation}
where $\phi_b$ is an irrelevant operator with scaling dimension $x_b>1$ on the boundary $y=0$.

We will assume the boundary conditions to be such that $\langle\phi_b(\tau)\rangle = 0$. This is natural if we demand conformal boundary conditions. Then the first-order correction vanishes. An important exception is when the perturbing boundary operator is the stress-energy tensor, a case that was treated in Ref.~\cite{SCLA}. Since this operator has a non-vanishing expectation value on $\rn$ it will give rise to a first-order correction to $S_A^{(n)}$, which was found to have the form $\ell^{-1}$. We therefore consider the second-order corrections to $F_{\rn}$ and $F$, denoted $\delta^2 F_{\rn}$ and $\delta^2 F$ respectively. They are given by
\begin{equation} \label{d2fn}
\delta^2 F_{\rn} = - \frac{\lambda^2}{2 \beta} 
 \int \mathrm{d} \tau'_1  \int \mathrm{d} \tau'_2  \, \langle \phi_b(\tau'_1)  \phi_b(\tau'_2) \rangle_{\mathcal{R}_n}
\end{equation} 
and
\begin{equation} \label{d2f}
\delta^2 F = - \frac{\lambda^2}{2 \beta} 
 \int \mathrm{d} \tau_1  \int \mathrm{d} \tau_2  \, \langle \phi_b(\tau_1)  \phi_b(\tau_2) \rangle_{\mathbb{C}^+} \, ,
\end{equation} 
respectively, where $\tau'_1$, $\tau'_2$ are boundary coordinates on $\rn$ and $\tau_1 = w(\tau'_1)$, $\tau_2 = w(\tau'_2)$ are boundary coordinates on $\mathbb{C}^+$. From Eq.~(\ref{trans}), we get
\begin{eqnarray}
\delta^2 F_{\rn} = - \frac{\lambda^2}{2 \beta} 
 \int \mathrm{d} \tau'_1  \int \mathrm{d} \tau'_2  \, \left| \frac{\mathrm{d}z}{\mathrm{d}w}  \right|_{z=\tau'_1}^{-x_b} \left| \frac{\mathrm{d}z}{\mathrm{d}w}  \right|_{z=\tau'_2}^{-x_b} \langle \phi_b(w(\tau'_1)) \phi_b(w(\tau'_2)) \rangle_{\mathbb{C^+}} \nonumber \\
 =- \frac{\lambda^2}{2 \beta} 
 \int \mathrm{d} \tau_1  \int \mathrm{d} \tau_2  \, \left| \frac{\mathrm{d}z}{\mathrm{d}w}  \right|_{w=\tau_1}^{1-x_b} \left| \frac{\mathrm{d}z}{\mathrm{d}w}  \right|_{w=\tau_2}^{1-x_b} \langle \phi_b(\tau_1) \phi_b(\tau_2) \rangle_{\mathbb{C^+}} \, . \label{d2fn2}
\end{eqnarray} 
We can now use the fact that $\delta^2 F_{\rn} - n \delta^2 F$ only depends on the ratio $\ell/\epsilon$, where $\epsilon$ is the short-distance cutoff of the theory, to extract its $\ell$-dependence. Since the action~(\ref{action}) is dimensionless, the coupling constant $\lambda$ goes as $\lambda \sim \epsilon^{x_b-1}$. Thus $\lambda^2 \sim \epsilon^{2x_b-2}$, and since $\mathrm{d}z / \mathrm{d}w \propto \ell$ the integral in Eq.~(\ref{d2fn2}) includes an overall factor of $(\ell/\epsilon)^{2-2x_b}$. However, there can also appear powers of $\ell/\epsilon$ coming from the need to regularize divergences in the integrals. 

In order to compare the two integrals in $\delta^2 F_{\rn} - n \delta^2 F$ it is convenient to rewrite $n \delta^2 F$ on the same form as $\delta^2 F_{\rn} $.
\begin{eqnarray}
n \delta^2 F &=& -\frac{\lambda^2}{2\beta} n \int \mathrm{d} \tau_1  \int \mathrm{d} \tau_2 \frac{1}{|\tau_1 - \tau_2|^{2x_b}} \nonumber \\& =&  -\frac{\lambda^2}{2\beta} n \int \mathrm{d} \tau_1  \int_{\epsilon}^{\infty} \mathrm{d} |\tau_1 - \tau_2| \frac{1}{|\tau_1 - \tau_2|^{2x_b}}\nonumber \\ &=&  \frac{\lambda^2}{2\beta}  n \int \mathrm{d} \tau_1 \nonumber \frac{\epsilon^{1-2x_b}}{1-2x_b} \, .
\end{eqnarray}
Rewriting this as an integral over the boundary of $\rn$, one gets
\begin{eqnarray}
n\delta^2 F &=& \frac{\lambda^2}{2\beta}   \int \mathrm{d} \tau'_1 \frac{\epsilon^{1-2x_b}}{1-2x_b} \nonumber \\ &=& \frac{\lambda^2}{2\beta}  \int \mathrm{d} \tau_1 \left|\frac{\mathrm{d}z}{\mathrm{d}w}\right|_{w=\tau_1}   \frac{\epsilon^{1-2x_b}}{1-2x_b} 
\end{eqnarray}
and then going back to writing this as a double integral over $\tau_1$ and $\tau_2$ gives
\begin{eqnarray}
n\delta^2 F &=& -\frac{\lambda^2}{2\beta} \int \mathrm{d} \tau_1  \int_{|\tau_1 - \tau_2| \geq \epsilon / |({\tiny \mathrm{d}}z / {\tiny \mathrm{d}}w)_{w=\tau_1}|} \mathrm{d} \tau_2 \left|\frac{ \mathrm{d}z}{\mathrm{d}w}\right|_{w=\tau_1}^{2-2x_b}   \frac{1}{|\tau_1  - \tau_2|^{2x_b}} \, . \nonumber \\
\end{eqnarray}
As $\epsilon \to 0$, we have
\begin{equation}
|( \mathrm{d}z /  \mathrm{d}w)_{w=\tau_1}| |\tau_1 - \tau_2| \geq \epsilon  \ \ \Leftrightarrow \ \ |\tau'_1 - \tau'_2| \geq \epsilon \, ,
\end{equation}
so that $\ff$ can be written as a single integral
\begin{eqnarray}
\ff &=& -\frac{\lambda^2}{2\beta}  \int \mathrm{d} \tau_1  \int \mathrm{d} \tau_2    \frac{\left|\frac{ \mathrm{d}z}{\mathrm{d}w}\right|_{w=\tau_1}^{1-x_b} \left|\frac{ \mathrm{d}z}{\mathrm{d}w}\right|_{w=\tau_2}^{1-x_b}  - \left|\frac{ \mathrm{d}z}{\mathrm{d}w}\right|_{w=\tau_1}^{2-2x_b}}{|\tau_1  - \tau_2|^{2x_b}} \nonumber \\
&=& \frac{\lambda^2}{4\beta}  \int \mathrm{d} \tau_1  \int \mathrm{d} \tau_2    \frac{\left( \left|\frac{ \mathrm{d}z}{\mathrm{d}w}\right|_{w=\tau_1}^{1-x_b} - \left|\frac{ \mathrm{d}z}{\mathrm{d}w}\right|_{w=\tau_2}^{1-x_b}  \right)^2 }{|\tau_1  - \tau_2|^{2x_b}} \label{int2}
\end{eqnarray}
with the cutoff $|( \mathrm{d}z /  \mathrm{d}w)_{w=\tau_1}| |\tau_1 - \tau_2| \geq \epsilon $. It follows from Eq.~(\ref{der}) that $|$d$z$/d$w|^{1-x_b}$ is analytic everywhere except at $w=\pm$i. In Ref.~\cite{CC3}, where the integrals are over $\cp$, this singularity at the branch point was an important ingredient in the analysis. But when we now consider a perturbing operator on the boundary, the only divergence in the integrand in (\ref{int2}) comes when $\tau_1 = \tau_2$. Expanding $|\mathrm{d}z /  \mathrm{d}w|^{1-x_b} \equiv f(w)$ around $w=\tau_2$, gives
\begin{eqnarray}
\hspace{-2cm} \ff &=& \frac{\lambda^2}{4\beta}  \int \mathrm{d} \tau_1  \int \mathrm{d} \tau_2    \frac{\left( f'(\tau_2) (\tau_1-\tau_2) + \frac{1}{2}  f''(\tau_2) (\tau_1-\tau_2)^2 + ...    \right)^2 }{|\tau_1  - \tau_2|^{2x_b}}  \nonumber \\
&=& \frac{\lambda^2}{4\beta}  \int \mathrm{d} \tau_1  \int \mathrm{d} \tau_2 \ [\   (f'(\tau_2))^2  |\tau_1-\tau_2|^{2-2x_b} \nonumber \\
&& \hspace{3cm} + f'(\tau_2)  f''(\tau_2)(\tau_1-\tau_2) |\tau_1-\tau_2|^{2-2x_b} + ...\   ]\,.  \label{int3} 
\end{eqnarray}
From this it follows that the leading divergence of the double integral in (\ref{int2}) goes as $\epsilon^{3-2x_b}$, i.e. it converges when $x_b < 3/2$. Then no regularization is needed, and the only $\ell$-dependence comes from d$z$/d$w \propto \ell$. Thus, when $x_b < 3/2$ 
\begin{equation}
\ff \sim (\ell/\epsilon)^{2-2x_b} \, ,
\end{equation} 
and consequently the leading corrections $\delta^2 S_A^{(n)}$ to the R{\'e}nyi entropies are of the form
\begin{equation} \label{result}
\delta^2 S_A^{(n)} \sim \ell^{2-2x_b} \, .
\end{equation} 
On the other hand, when $x_b > 3/2$, the cutoff in the integral (\ref{int2}) must be kept, so that
\begin{equation}
\ff \sim (\ell/\epsilon)^{2-2x_b} (\ell/\epsilon)^{2x_b-3} = (\ell/\epsilon)^{-1}\, ,
\end{equation} 
and then $\delta^2 S_A^{(n)} \sim \ell^{-1}$ for all $x_b > 3/2$. Note that this is of the same form as the first-order correction from the stress-energy tensor. 

When $x_b=3/2$, it follows from Eq.~(\ref{int3}) that the integral (\ref{int2}) diverges logarithmically, hence 
\begin{equation} \label{result}
\delta^2 S_A^{(n)} \sim \ell^{-1} \log \ell \, .
\end{equation} 
These deviations from the $\sim \ell^{2-2x_b}$ behavior of the correction when $x_b \geq 3/2$ goes beyond what would be expected from standard finite-size scaling arguments.

\section{Scaling corrections from marginal boundary operators}

When perturbing with a marginal boundary operator one cannot simply put $x_b=1$ in Eq.~(\ref{int2}) and conclude that the second-order corrections to scaling of the R{\'e}nyi entropies vanish, since $|$d$z$/d$w|$ diverges when $|z|\to \infty$. However, it can be checked that there is no need to regularize the integral because of this. We therefore conclude that the second-order corrections will be $\ell$-independent when $x_b=1$. 

Instead of going to the higher-order integrals in the perturbation series of $F_{\rn} - nF$ to find the leading $\ell$-dependence of the corrections we will make use of the $g$-theorem~\cite{AL}, analogously to how Cardy and Calabrese~\cite{CC3} use the $c$-theorem in the marginal bulk case. 

The boundary entropy $s^A$ of Eq.~(\ref{S_A3}) is governed by the "gradient formula" of Friedan and Konechny~\cite{FK1} which in our case takes the simple form
\begin{equation} \label{gradient}
\frac{\partial s^A}{\partial \lambda} = - \beta(\lambda) \, ,
\end{equation} 
where $\beta$ is the renormalization-group beta function given by~\cite{cardy}
\begin{equation} \label{beta}
-\beta(\lambda) = \ell \frac{\mathrm{d}\lambda}{\mathrm{d}\ell} = (1-x_b) \lambda - \pi b \lambda^2 + \mathcal{O}(\lambda^3) \, .
\end{equation} 
In the marginally irrelevant case, i.e. with $x_b=1$ and $\lambda/b>0$, this gives
\begin{equation} \label{betamarg}
\ell \frac{\mathrm{d}\lambda}{\mathrm{d}\ell} =  - \pi b \lambda^2 + \mathcal{O}(\lambda^3)\, ,
\end{equation} 
with the asymptotic large-$\ell$ solution given by
\begin{equation} \label{lambda}
\lambda(\ell) \sim \frac{1}{\pi b \log (\ell/\epsilon)}. 
\end{equation} 
Now, as Eq.~(\ref{gradient}) becomes $\partial s^A / \partial \lambda = - \pi b \lambda^2 + \mathcal{O}(\lambda^3)$ when $x_b=1$, we have
\begin{equation} \label{sA}
s^A = \mathrm{const.} -\frac{\pi b}{3}  \lambda^3 + \mathcal{O}(\lambda^4) \sim \log g^A -\frac{1}{3 \pi^2 b^2 \log^3 (\ell/\epsilon)} \, .
\end{equation}
Thus the leading correction to the R{\'e}nyi entropies goes as $(\log \ell)^{-3}$.

\section{Conclusions}

The result that the leading second-order corrections to the R{\'e}nyi entropies $S_A^{(n)}$ are of the form $\ell^{2-2x_b}$ when perturbing with an irrelevant boundary operator with scaling dimension $x_b<3/2$ holds regardless of the value of $n$. This is the result anticipated from finite-size scaling. When $x_b > 3/2$ the leading corrections will be of the same form as those from the stress-energy tensor, i.e. $\ell^{-1}$. When $x_b=3/2$ there is also a multiplicative logarithmic contribution to the leading correction which then goes as $\ell^{-1}\log \ell$. A marginally irrelevant boundary perturbation gives a correction $\sim (\log \ell )^{-3}$. Thus there are no unusual $n$-dependent corrections to scaling of the R\'enyi entropies from boundary operators, as opposed to bulk perturbations where unusual corrections to scaling can occur. These unusual corrections originate from the part of the surface integral where the bulk operator approaches the branch point created by the Riemann surface construction~\cite{CC3}. However, when the perturbing field is \textit{on} the boundary it never comes close to this singularity.

In Ref.~\cite{SCLA} it was found that the first-order correction from the stress-energy tensor on the boundary is $\sim \ell^{-1}$. It should be noted that this operator is generically present, and gives a correction to the entanglement entropy of the same form as that from a boundary operator with scaling dimension $x_b >3/2$.

We also note that the leading corrections of the form $\ell^{2-2x_b}$ from a boundary perturbation are similar to what one can get when perturbing with a bulk operator in the presence of a boundary. In Ref.~\cite{CC3} it was found that this can give corrections of the form $\ell^{2-x}$, where $x$ is the bulk scaling dimension, but also unusual corrections which can dominate. This is therefore a very different situation compared to having the perturbing field on the boundary.

\section*{Acknowledgments}
We wish to thank Pasquale Calabrese for helpful discussions. We also acknowledge NORDITA for hospitality during the completion of
this work. This research was supported by
the Swedish Research Council under Grant No. VR-2008-4358.

\end{document}